\documentstyle[aps,prl,twocolumn,floats,psfig,epsf]{revtex}

\def\spose#1{\hbox to 0pt{#1\hss}}
\def\ltapprox{\mathrel{\spose{\lower 3pt\hbox{$\mathchar"218$}}
 \raise 2.0pt\hbox{$\mathchar"13C$}}}
\def\gtapprox{\mathrel{\spose{\lower 3pt\hbox{$\mathchar"218$}}
 \raise 2.0pt\hbox{$\mathchar"13E$}}}

\begin{document}
\draft
\twocolumn[\hsize\textwidth\columnwidth\hsize\csname@twocolumnfalse\endcsname
\title{
$k$-string tensions in SU($N$) gauge theories
}
\author{Luigi Del Debbio$^a$, Haralambos Panagopoulos$^b$, Paolo Rossi$^a$,
and Ettore Vicari$^a$}
\address{$^a$ Dipartimento di Fisica dell'Universit\`a di Pisa
and I.N.F.N., I-56127 Pisa, Italy
}
\address{$^b$ Department of Physics, University of Cyprus,
                      Nicosia CY-1678, Cyprus
\\
{\bf e-mail: \rm 
{\tt ldd@df.unipi.it},
{\tt haris@ucy.ac.cy},
{\tt rossi@df.unipi.it}, 
{\tt vicari@df.unipi.it}
}}

\date{\today}
\maketitle

\begin{abstract}
In the context of four-dimensional SU($N$) gauge theories, we study the spectrum
of the confining strings. We compute, for the 
SU(6) gauge theory formulated on a lattice, the three independent string tensions 
$\sigma_k$ related to sources with $Z_N$ charge $k=1,2,3$,
using Monte Carlo simulations.
Our results, whose uncertainty is approximately 2\% for $k=2$ and
$4\%$ for $k=3$, are consistent with the sine formula 
$\sigma_k/\sigma = \sin k \case{\pi}{N} / \sin \case{\pi}{N}$
for the ratio between $\sigma_k$ and the 
standard string tension $\sigma$, 
and show deviations from the Casimir scaling.

The sine formula is known to emerge
in supersymmetric SU($N$) gauge theories and in M-theory.
We comment on an analogous behavior exhibited by 
two-dimensional SU($N$) $\times$ SU($N$)  chiral models.
\end{abstract}


]

Quantum Chromodynamics
is a nonabelian gauge theory based on the gauge group SU(3). The
mechanisms underlying many of its
fundamental properties, such as confinement, chiral symmetry, topological effects
and the axial anomaly, are under active investigation; they are being studied
by different approaches, including numerical simulations of the theory
formulated on the lattice, 
several models of the vacuum,
as well as some recent proposals derived from
M-theory and AdS/CFT.  
Many features of QCD can be better understood by extending 
the study to SU($N$) gauge theories with $N$ larger than three
and in particular by examining the large-$N$ limit.

Four-dimensional gauge theories exhibit confinement, i.e. 
static sources in the fundamental representation develop
a linear potential  characterized by a string tension $\sigma$.
As pointed out in many studies, it is important to investigate
the behavior of the system in the presence of static sources in
representations higher than the fundamental one. 
This may provide useful hints on the mechanism
responsible for confinement, helping to identify the most appropriate
models of the QCD vacuum and to select among the various confinement
hypotheses.
Among the latter, the so-called Casimir scaling hypothesis
for the potential between heavy-quark sources in different
representations has attracted much interest (see e.g. the
recent publications~\cite{SS-00,KITS-01,Bali-00,Deldar-00,FGO-98}).

SU($N$) gauge theories confine by means of chromoelectric flux tubes carrying charge in the
center $Z_N$ of the gauge group. A chromoelectric source of charge $k$
with respect to  $Z_N$ is confined by a $k$-string with string tension
$\sigma_k$ ($\sigma_1\equiv \sigma$ is the string tension related to 
the fundamental representation). 
If $\sigma_k < k\, \sigma$, then a string with
charge $k$ is stable against decay to $k$ strings of charge one.
Charge conjugation implies 
$\sigma_k=\sigma_{N-k}$. Therefore $SU(3)$ has
only one independent string tension determining the large distance
behavior of the potential for $k\ne 0$.
One must consider larger values of $N$
to look for distinct $k$-strings.

As pointed out in Ref.~\cite{Strassler-98}, it is interesting
to compare the $k$-string tension ratios 
\begin{equation}
R(k,N) \equiv  {\sigma_k\over \sigma}
\end{equation}
in different theories. The idea is that such ratios may reveal
a universal behavior within  a large class of models characterized by
SU($N$) symmetry, such as SU($N$) gauge theories and their
supersymmetric extensions. 
It has been noted that
stable $k$-strings are related to the totally antisymmetric 
representations of rank $k$, and that 
in various realizations of supersymmetric SU($N$) gauge theories
$R(k,N)$ satisfies the sine formula $R(k,N)=S(k,N)$ where
\begin{equation}
S(k,N) \equiv { \sin (k\pi/N) \over  \sin (\pi /N)}\,.
\label{sinf}
\end{equation}
$R(k,N)$ has been computed 
for  the ${\cal N} = 2$ supersymmetric
SU($N$) gauge theory softly broken  to ${\cal N} = 1$
\cite{DS-95,HSZ-98}, obtaining Eq. (\ref{sinf}).
The same result is found also in the context of M-theory,
and extended to the case of large breaking of the ${\cal N}=2$
supersymmetric theory \cite{HSZ-98}.
An interesting question is
whether the sine formula holds in nonsupersymmetric 
SU($N$) gauge theories.
The M-theory approach to nonsupersymmetric QCD, although
it is still at a rather speculative stage, 
suggests that it may be so \cite{Witten-97,HSZ-98}. However,
as discussed in Refs.~\cite{HSZ-98,Strassler-98}, 
corrections from various sources cannot be excluded,
so that this prediction cannot be considered robust.

Another interesting and suggestive hypothesis is that the $k$-string
tension ratio satisfies 
the so-called Casimir scaling law \cite{AOP-84}, i.e. 
$R(k,N)=C(k,N)$ where
\begin{equation}
C(k,N) \equiv {k(N-k)\over N-1}
\label{casf}
\end{equation}
is the ratio between the values of the quadratic Casimir
operators in the rank-$k$ antisymmetric and in the 
fundamental representations.
The Casimir ratio is satisfied on the one hand  by
the strong-coupling limit of the lattice Hamiltonian formulation
of SU($N$) gauge theories 
\cite{KS-75}, and on the other hand 
by the small-distance behavior of the potential between two static charges in different
representations, as shown by perturbation theory up to two
loops~\cite{pert-pot}.
Interest in Casimir scaling was recently
revived~\cite{SS-00,KITS-01,Bali-00,Deldar-00,FGO-98}; it has been
triggered by  numerical studies of SU(3) lattice gauge
theory \cite{Bali-00,Deldar-00}, which indicate that Monte Carlo 
data for the potential between charges in different
representions are consistent with Casimir scaling
up to a relatively large distance,
$r \approx 1 {\rm fm}$. 

The Casimir scaling law holds exactly in two-dimensional QCD. 
In higher dimensions no strong arguments exist in favor of
a mechanism preserving Casimir scaling across the roughening
transition, from strong to weak coupling; nor from small
distance (essentially perturbative, characterized by a 
Coulombic potential) to large distance
(characterized by a string tension for sources carrying $Z_N$
charge).
We have shown explicitly~\cite{noiinprep} that Casimir scaling
does not survive the next-to-leading order calculation of
the ratios $R(k,N)$ in the strong-coupling lattice Hamiltonian
approach.

It is interesting to note that the sine formula (\ref{sinf})
emerges also in the context of the two-dimensional SU($N$)$\times$ SU($N$)
chiral models.
As amply discussed in the literature (see 
e.g. Refs.~\cite{Polyakov-88,RCV-98} and references therein),
$d$-dimensional chiral models and $2d$-dimensional lattice gauge theories
manifest deep  analogies in the continuum and on the lattice. 
Indeed, one may establish the following correspondence table for the
lattice formulations:
\begin{eqnarray*}
\begin{tabular}{c@{\quad}c}
{\bf Chiral Models} & {\bf Gauge Models} \\
site, link & link, plaquette \\
loop & surface \\
length & area \\
mass $M$ & string tension $\sigma$ \\
two-point correlation & Wilson loop \\
\end{tabular}
\end{eqnarray*}
One may also add to this table 
the bound state masses $M_k$ 
of chiral models and the $k$-string tensions $\sigma_k$
of gauge theories.
In particular, in the case $d=1$ the
relation is exact, and one can  prove that Casimir scaling
holds for the masses of the bound states.
In analogy to four-dimensional SU($N$) gauge theories,
in two-dimensional SU($N$)$\times$ SU($N$) chiral models
the Casimir scaling law holds for
the strong-coupling limit of the corresponding
lattice Hamiltonians
(but it is not satisfied by the corrections) 
and for the small-distance behavior of 
the correlation functions related to different  representations.
On the other hand,
the exact S-matrix \cite{chiralS}, derived using the existence of an infinite number
of conservation laws and 
Bethe Ansatz methods, dictates that 
all bound states belong to the rank-$k$ antisymmetric
representations and satisfy the sine formula, 
$M_k=M \sin ( k \case{\pi}{N}) / \sin (\case{\pi}{N})$, 
where $M_k$ is the mass of the $k$-particle bound state.
The question arises again: does this result extend to four-dimensional
SU($N$) gauge theories?

This issue  can be investigated
numerically using the lattice formulation of SU($N$) gauge theories.
Recent numerical results for $R(2,N)$, obtained for $N=4,5$
\cite{LT-01,WO-01}, show that $R(2,N)<2$\,; thus, $\sigma_2 < 2\sigma$, 
indicating that flux tubes attract each other.
However, the error estimates on $R(2,N)$ do not
allow to exclude any of the two above-mentioned
hypotheses. Indeed the sine and Casimir formulas give numerically
close predictions for $k=2$, so that high accuracy is necessary to distinguish them.

In this work we further investigate the spectrum of the string states.
We present results from Monte Carlo (MC) simulations of the
four-dimensional SU(6) lattice gauge theory using the Wilson
formulation.  
For $N=6$ there are two nontrivial $k$-string tensions besides the fundamental one,
thus providing a stringent test of the various ideas discussed above.
We anticipate here our final results for the two independent 
$k$-string tension ratios:
\begin{eqnarray}
R(2,6) &=& 1.72  \pm 0.03 ,\label{finest2}\\
R(3,6) &=& 1.99 \pm 0.07. \label{finest3}
\end{eqnarray}
They are both consistent with the predictions of the sine
formula (\ref{sinf}), which are $S(2,6) = 1.732...$ and $S(3,6)=2$, respectively.
On the other hand, our results show deviations from  Casimir scaling;
the predictions in this case,  
$C(2,6)=1.6$ and $C(3,6)=1.8$, 
are off by approximately four and three error bars, respectively.

\begin{table}[htp]
\caption{MC results for the $k$-string tensions.}
\label{tab:su6data}
\begin{tabular}{crrrl}
\multicolumn{1}{c}{$\gamma$}&
\multicolumn{1}{c}{lattice}&
\multicolumn{1}{c}{$a \sqrt{\sigma}$}&
\multicolumn{1}{c}{$\sigma_2/\sigma$}&
\multicolumn{1}{c}{$\sigma_3/\sigma$}\\
\tableline \hline
0.342   & $ 8^3\times 20$   & 0.3151(6) & 1.65(2)  & 1.91(3)  \\
        & $ 12^3\times 24$  & 0.3239(8) & 1.66(3)  & 1.91(9)  \\

0.344   & $ 12^3\times 24$  & 0.2973(5) &  1.73(2) & 1.95(5)  \\

0.348   & $ 10^3\times 20$  & 0.2534(6) &  1.73(3) & 2.08(10)  \\
        & $ 12^3\times 24$  & 0.2535(6) &  1.71(4) & 2.06(11)  \\

0.350   & $ 12^3\times 24$  & 0.2380(6) &  1.72(3) &  1.95(9) \\

0.354   & $ 12^3\times 24$  & 0.2103(5) &  1.73(3) &   2.04(6) \\
\end{tabular}
\end{table}

In our simulations we employed the Cabibbo-Marinari algorithm
\cite{CM-82} to upgrade SU($N$) matrices by updating their SU(2)
subgroups (we selected 15 subgroups). This was done by alternating
microcanonical over-relaxation and heat-bath steps, typically in a 4:1
ratio. Table~\ref{tab:su6data} contains some information on our
MC runs: The coupling values \cite{gamma} $\gamma\equiv \beta/(2 N^2)$,
lattice sizes, and the results for the $k$-string
tensions. The number of sweeps per run was
typically above 500k, and measurements were taken every 10-20 sweeps.
The values of $\gamma$ were chosen to lie beyond the first order phase transition
which occurs in the Wilson formulation of SU($N$) gauge theories for $N$ sufficiently
large (see e.g. Refs.~\cite{phasetr,noiinprep}).
In order to determine the value $\gamma_c$
where the first order  transition occurs, we 
performed simulations starting from hot and cold configurations
to display hysteresis, and from mixed-phase
configurations, obtaining the estimate $\gamma_c=0.3389(4)$.
We used asymmetric lattices ($L^3\times T$) with a larger time size.
For some values of $\gamma$
we performed simulations for two lattice sizes, to check for
finite size effects.  The lattice sizes $L$ were chosen so that
$L\sqrt{\sigma}\gtapprox 2.5$, and for most of them
$L\sqrt{\sigma}\approx 3$. 
This requirement ensures that finite size effects on $k$-string ratios
are negligible, as can be seen by comparison of the results for different
sizes (see also Refs.~\cite{LT-01}).
Further confirmation comes from
preliminary results ($\approx$ 200k sweeps)
on a $16^3\times 32$ for $\gamma=0.350$.

In our simulations we have also measured the topological charge $Q$, by
a cooling technique. A severe form of critical slowing down is
observed in this case:
The autocorrelation time $\tau_Q$
for $Q$ appears to increase exponentially,
$\tau_Q\propto \exp (c/\sigma^{1/2})$  with $c\approx 2.4$.
As a consequence, the run for the largest value of 
$\gamma$ considered, $\gamma=0.354$,
did not correctly sample $Q$, presumably because 
it was not sufficiently long ($\approx$ 300k sweeps). 
This dramatic effect was not observed 
in the correlations used to determine the $k$-string tensions
(a blocking analysis
did not show significant time correlations in measurements taken every
10-20 sweeps),  
suggesting an approximate
decoupling between the topological and nontopological modes.  
This suggestion is also supported by the fact that string tension
results for $\gamma=0.354$,
extracted from a simulation which did not sample correctly $Q$, 
turn out to be in agreement with those for smaller
$\gamma$, for which $Q$ was sampled correctly.
We note incidentally that this phenomenon has already been observed 
in simulations of two-dimensional CP$^{N-1}$ models \cite{CPN-Q}.

The $k$-string tensions are extracted from the large time behavior of
correlators of strings in the antisymmetric representations, closed
through periodic boundary conditions (see e.g.
Refs.\cite{DSST-85,LT-01}):
\begin{equation}
C_r(t) = \sum_{x_1,x_2} \langle \chi_r [ P(0;0) ]\; 
\chi_r [ P(x_1,x_2;t) ] \rangle
\end{equation}
where $P(x_1,x_2;t) = \Pi_{x_3} U_3(x_1,x_2,x_3;t)$,
$U({\bf x}; t)$ are the usual link variables, and $\chi_r$ is the
character of the representation $r$. In particular, 
$\chi_{f}[ P ] = {\rm Tr}\, P$ for the fundamental
representation, and 
\begin{eqnarray}
&&\chi_{k=2}[ P ] = {\rm Tr}\, P^2 - \left( {\rm Tr} \,P\right)^2,\\
&&\chi_{k=3}[ P ] = 2 {\rm Tr}\, P^3 - 3 {\rm Tr}\, P^2 \; {\rm Tr}\,P +
\left( {\rm Tr} \,P\right)^3
\end{eqnarray} 
for the antisymmetric representations  of rank $k=2$
and $k=3$, respectively.

\begin{figure}
\vspace*{0truecm} \hspace*{-0.2cm}
\centerline{\psfig{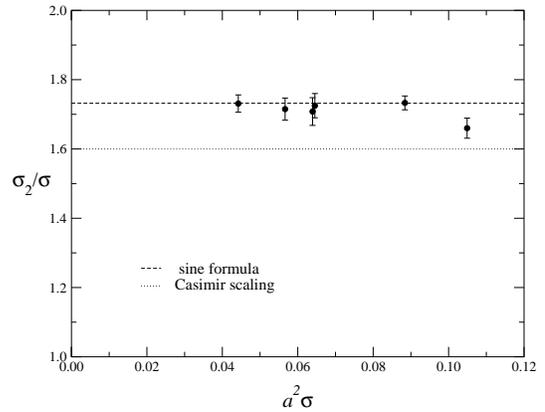}}
\vspace*{0.cm}
\caption{The scaling ratio $R(2,6)$ as a function of $a^2 \sigma$.
}
\label{fig:R26}
\end{figure}

These correlators decay exponentially as $\exp (- m_k t)$ where $m_k$ is the
mass of the lightest state in the corresponding representation.
For a $k$-loop of size $L$, the $k$-string tension is obtained
using the relation \cite{DSST-85}
\begin{equation}
m_k = \sigma_k L - {\pi\over 3 L}.
\label{mks}
\end{equation}
The last term in Eq.~(\ref{mks}) is conjectured to be a universal correction, and
it is related to the universal critical behavior of the flux
excitations described by a free bosonic string
\cite{LSW-80,DSST-85}.  
In order to improve the efficiency of the measurements we used
smearing and blocking procedures (see e.g. Refs.~\cite{smearing}) to
construct new operators with a better overlap with the lightest 
state.
The masses $m_k$ were obtained by fitting the time dependence of the
correlations.
The fitting range is source of systematic error; we have checked
that all reasonable choices of this range yield consistent results
within the quoted errors. 
More details on the Monte Carlo simulations and the analysis of the
data will be reported elsewhere \cite{noiinprep}.

\begin{figure}
\vspace*{0truecm} \hspace*{-0.2cm}
\centerline{\psfig{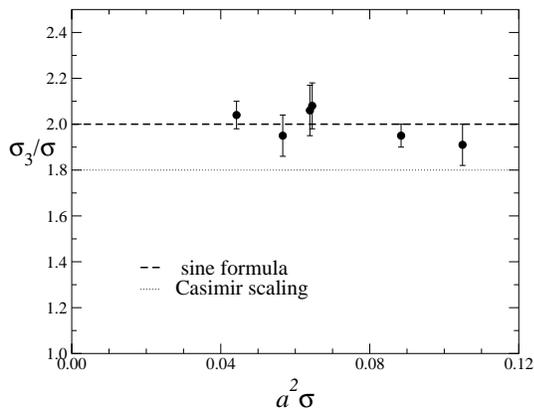}}
\vspace*{0.cm}
\caption{The scaling ratio $R(3,6)$ as a function of $a^2 \sigma$.
}
\label{fig:R36}
\end{figure}

The results for the ratios $R(2,6)$ and $R(3,6)$ 
are presented in Table~\ref{tab:su6data}, and plotted
in Figs.~\ref{fig:R26} and \ref{fig:R36} versus $a^2\sigma$
to evidentiate possible scaling corrections,
for the case $k=2$ and
$k=3$ respectively, together with the sine
formula (\ref{sinf}) and the Casimir scaling predictions.
The ratio $R(2,6)$ shows good scaling for $\gamma\ge 0.344$.
Scaling deviations are observed only for $\gamma=0.342$, and 
this may be due to the vicinity of the phase transition.  
The data for $\gamma\ge 0.344$ are consistent with a constant, thus
we do not attempt to fit the dependence of
our result on the lattice spacing $a$. Our final value for $R(2,6)$
is obtained by combining the results at $\gamma=0.348$ 
(for the largest lattice) and
$\gamma=0.350$. The error we report is given by the typical error of each single
point. 
Of course, this estimate assumes that the scaling corrections
are small and negligible for $a^2\sigma\simeq 0.05$.
The data for smaller $\gamma$, and in particular the one for
$\gamma=0.344$, are essentially used to check this fact.
They suggest that the  scaling corrections are at most of the same
size of
the error we report. 
MC runs at the largest value of
$\gamma$, i.e. $\gamma=0.354$,  show the aforementioned decoupling of the string tensions
from the topological degrees of freedom; however, given the poor
sampling of the topological charge in those runs, we do not include them
in the final estimate of the string tension ratio. Similar comments
apply to the $R(3,6)$ ratio. 

We have also explored correlators in the symmetric rank-2
representation, finding no evidence for stable bound states, in accordance
with general arguments and with the spectrum of chiral models.

Our final estimates have been reported in Eqs.~(\ref{finest2})
and (\ref{finest3}).
They show deviations from Casimir scaling \cite{scalcorr}.
It is worthwhile to
emphasize that such corrections are to be expected, as mentioned
previously.
This fact is further confirmed by the computation of the ratio
$\sigma_k/\sigma$ to $O(g^{-8})$ in the strong-coupling expansion
of the lattice Hamiltonian formulation of $d$-dimensional SU($N$)
gauge theories. 
We obtained \cite{noiinprep}
\begin{equation}
{\sigma_k\over \sigma} = 
{k(N-k)\over N-1} \left[ 1 + {(d-2) f(k,N) \over (g^2 N)^4} 
+ ... \right]
\end{equation}
where $f(k,N)$ is explicitly $k$-dependent. 
In particular $f(2,N) = \case{6}{N} + O(\case{1}{N^2})$.

In conclusion, we claim that 
our numerical results for the four-dimensional SU(6) gauge 
theory are consistent with the sine formula, and the universality
hypothesis that is behind it.  
Of course, they do not prove that
it holds exactly. But they put a stringent
bound on the size of the possible corrections.
On the other hand, our results
show a clear evidence of deviations from the  Casimir scaling.
This fact  should be relevant for the
recent debate on confiniment 
models, such as those discussed in Refs. \cite{SS-00,KITS-01,FGO-98}.
However, Casimir scaling  may still be considered as a reasonable
approximation, since the largest deviations we observed
were about 10\%.

One last remark regards the large-$N$ behavior of the
sine formula:
$S(k,\infty) = k + O\left( {1/ N^2}\right)$.
In this respect the sine formula is peculiar
because there are no a priori reasons for the $k$-string tension ratio
to be even in $1/N$. The same observation applies to the
two-dimensional SU($N$)$\times$ SU($N$) chiral models, but there we
know that the sine formula holds and it comes from the structure of
the S-matrix, which is essentially determined by the existence of an
infinite number of conservation laws.

\medskip
{\bf Acknowledgments.} We thank M. Campostrini, K. Konishi, 
S. Lelli, B. Lucini, M. Maggiore,
A. Pelissetto, and M.
Teper for useful and interesting discussions, and 
M. Davini for his indispensable technical support.


\begin{references}
\bibitem{gamma}
In order to compare Monte Carlo results for various values of $N$, it
is useful to introduce the rescaled coupling
$\gamma \equiv {1/ (g_0^2 N)} = {\beta / (2 N^2)}$, which is kept fixed
when taking the large-$N$ limit of the lattice theory.
\bibitem{scalcorr}
In order to further check the robustness of our statement concerning
the deviation from the Casimir formula, we have analyzed
our data using also fits which allow for scaling corrections.
Of course, for this purpose one must consider all data,
including the one at $\gamma=0.342$, since it is the
only one showing apparent scaling violations.
As suggested in Ref.~\cite{LT-01}, one may fit the data 
to the linear behavior $A+B\sigma$.
The results are $R(2,6)=1.77(5)$ and $R(2,6)=1.79(6)$ respectively
with and without the point at the largest value of $\gamma$, i.e.
$\gamma=0.354$. 
Therefore, if one wants to be more cautious in treating the 
systematic error due to scaling corrections, one may take  
into account the linear fits, 
and arrive at  
$R(2,6) = 1.72\pm 0.03\;^{+0.07}_{-0.00}\;$,
which covers all the results obtained above.
This conservative estimate is still not consistent with 
the Casimir formula. Thus, our conclusions are fully justified even 
by this overly cautious analysis.
\end{references}
\end{document}